# Unravelling the intrinsic and robust nature of van Hove singularities in twisted bilayer graphene

I. Brihuega (1), P. Mallet (2), H. González-Herrero (1), G. Trambly de Laissardière (3), M. M. Ugeda (1), L. Magaud (2), J.M. Gómez-Rodríguez (1), F. Ynduráin (1), and J.-Y. Veuillen(2,*)

(1): Dept. Física de la Materia Condensada, Universidad Autónoma de Madrid, E-28049 Madrid, Spain
(2): Institut Néel, CNRS-UJF, BP 166, F-38042 Grenoble, France
(3): Laboratoire de Physique Théorique et Modélisation, Université de Cergy-Pontoise-CNRS, F-95302 Cergy-Pontoise, France

Abstract:

Extensive scanning tunnelling microscopy and spectroscopy experiments complemented by first principles and parameterized tight binding calculations provide a clear answer to the existence, origin and robustness of van Hove singularities (vHs) in twisted graphene layers. Our results are conclusive: vHs due to interlayer coupling are ubiquitously present in a broad range (from 1° to 10°) of rotation angles in our graphene on 6H-SiC(000-1) samples. From the variation of the energy separation of the vHs with rotation angle we are able to recover the Fermi velocity of a graphene monolayer as well as the strength of the interlayer interaction. The robustness of the vHs is assessed both by experiments, which show that they survive in the presence of a third graphene layer, and calculations, which test the role of the periodic modulation and absolute value of the interlayer distance. Finally, we clarify the origin of the related moiré corrugation detected in the STM images.

Soon after the discovery of the unique electronic properties of graphene [1-3], suggestions were made for engineering the band structure of this material. It has been proposed that periodic potentials with wavelengths in the nanometre range could lead to anisotropic renormalization of the velocity of low energy charge carriers [4] or to the generation of new massless Dirac fermions [5]. Experimental works intended for verifying these theoretical predictions were recently reported [6-8], where the periodic perturbation was generated either by a lattice mismatch with the supporting material or by a self-organized array of clusters. An alternative route for modifying graphene's band structure would be to exploit a rotation between stacked graphene layers [9]. According to calculations, for large angles

($\theta \geq 15°$) the low energy band structure of graphene should be preserved [10-12]. For intermediate angles ($1° \leq \theta \leq 15°$), it is predicted that while the linear dispersion persists in the vicinity of the Dirac points of both layers, the band velocity is depressed and two saddle points appear in the band structure, giving rise to two logarithmic van Hove singularities (vHs) in the density of states (DOS) [9, 13-18] . For smaller angles ($\theta \leq 1°$) weakly dispersive bands appear at low energy [19, 20] with sharp DOS peaks very close to the Dirac point [17, 18].

Twisted graphene layers are commonly found on different substrates, like metals [13, 21, 22], the C-face of SiC [23-25] or graphite surfaces [26, 27]. Transfer techniques yielding large domains of twisted bilayers over a macroscopic sample [28] and quantitative, fast, Raman characterization tools [29, 30] have been recently proposed. However, despite rotated graphene layers are readily available and a number of measurements have confirmed that large twist angles lead to an electronic decoupling of stacked graphene layers [3, 11, 24, 31-34], few experiments tackle the electronic properties for sufficiently small angles ($\theta < 15°$) [13, 35, 36]. In particular, recent scanning tunnelling microscopy and spectroscopy (STM/STS) studies have demonstrated the renormalization of the band velocity [35] and the appearance of van Hove singularities in the local DOS (LDOS) [13] of three twisted layers configurations, one measured at the graphite surface and two on few layers graphene (FLG) grown on Ni. At variance, a careful angle resolved photoemission spectroscopy investigation of FLG on SiC-C face did detect neither a Van Hove singularity nor any significant change in the Fermi velocity in the range $1° < \theta < 10°$, and suggests major problems in our current understanding of twisted graphene layers [36].

Although twisted bilayers with small rotation angles (typically $\theta \leq 10°$) appear as a fascinating field of development for the physics of graphene [22, 37, 38], uncertainties about the mechanism of interlayer interaction remain [36, 39]. Here we show that graphene layers rotated between 1° and 10° present singularities in the LDOS. Our numerical simulations confirm that they arise from a partial gap opening at the crossing points of Dirac cones from neighbouring layers, and correspond to logarithmic vHs generated by the interlayer coupling. The vHs are found to be robust against deviations from the ideal twisted bilayer model both in the simulations and in the experiments on the real samples. We find that, as a result of the interlayer rotation, the graphene layers are, in addition, corrugated. The nature and effect of such corrugation in the vHs has been understood by a quantitative comparison with density functional theory (DFT) calculations.

As experimental realization of twisted layers we have chosen a 5 layers thick graphene grown in ultra-high vacuum on a 6H-SiC(000-1) substrate following the procedure described in Ref. [23]. STM images [40] reveal coherent domains with lateral size around 100 nm. Within these domains, the rotations between the surface graphene layers give rise to superstructures with period P in the nanometer range, which are identified as Moiré Patterns (MP) [23, 25, 26, 27]. The STM/STS experiments have been made in a separate UHV system using a home-made low temperature (5 K) STM [41]. Conductance spectra were taken using a lock-in technique, with an ac voltage (frequency: 830 Hz, amplitude: 1-10 mV rms) added to the dc sample bias. The sample was perfectly regenerated in situ by high temperature annealing [42, 43]. The data were analyzed using the WSxM software [44].

To settle the origin of the structures seen in STS data we have performed band structure and DOS calculations for twisted bilayers using two different approaches. The first one is a first principles DFT calculation based on the SIESTA code [45], including van der Waals interactions. This allows testing the roles of the atomic relaxation and corrugation, the interlayer distance, the importance of the van der Waals forces, etc [40]. The second one is a tight binding (TB) approach [14] which allows handling very large supercells.

The main experimental results of this report are summarized in figure 1. Fig. 1(c) shows STM images of 4 single period (simple) MP corresponding to rotations angles ranging from 9.6° to 1.4°. The rotation angle $\theta$ between neighbouring carbon planes can be obtained from the value of the moiré period P, using $\sin(\theta/2)=0.123/P(nm)$ [23, 27] and the drawing on Fig. 1(a) illustrates the MP for a 9.6° rotation angle. As depicted in Fig 1(b), the Brillouin Zones of the graphene layers are equally rotated by $\theta$. Thus, the Dirac cones of each layer are now centered in different points of the reciprocal space $K_1$ and $K_2$. The cones merge into two saddle points at energies $\pm E_{VHs}$ from the Dirac point, leading to vHs which generate peaks in the DOS. LDOS spectra taken on the MP in Fig. 1(c) are shown in Fig. 1(d). Each spectrum displays two peaks, indicated by arrows, identified as vHs by our theoretical calculations. They are approximately symmetric with respect to the Fermi level $E_F$ and evolve towards lower bias for decreasing rotation angle. We have measured the energy separation of the vHs as a function of the rotation angle $\theta$ for a large number ($\approx$30) of simple MP and the result is shown as filled circles in Fig. 1(e). We also plot as crosses the only three other data that had been reported till date, measured for small rotation angles ($\theta$<4°) in different substrates, which all fall at compatible energies [13]. Finally, as quoted previously [13], our spectra show that for small angles ($\theta$<3.5°) the peaks are localized in the bright regions of the MP, see the upper two spectra in Fig. 1(d) for θ=1.4°, a localization which disappears for larger angles, where vHs are completely homogenous across the MP [13, 40].

The DOS of twisted bilayers computed for rotation angles 1°<θ<10° are shown in figure 2. As in our experimental data, both ab-initio (Fig. 2(a)) and TB (Fig. 2(b)) approaches show two main peaks, which shift towards larger energy with increasing angle. Band structure calculations [14, 17, 40] show that these peaks are associated with the avoided crossing of the bands of the two layers along the line connecting $K_1$ and $K_2$, as illustrated in Fig 1(b). The calculations thus demonstrate that the peaks correspond to the vHs singularity described for smaller rotation angles [13]. The TB calculations of the LDOS in Fig. 2(b) show that the vHs tend to localize in the regions of the MP with AA stacking at small rotation angle [14]. This is again in agreement with experimental data (Fig. 1(b) and Ref. [13]), since these regions correspond to the bright areas in the STM images of the MPs, as shown below.

According to the expression derived from the continuum model [9, 13, 18], the energy separation of the vHs, $\Delta E_{VHs}$, follows: $\Delta E_{VHs}=2\hbar.v_F.\Gamma K.\sin(\theta/2)-2t_\theta$ (1), where $v_F$ is the Fermi velocity for monolayer graphene, $\Gamma K=1.703$ Å$^{-1}$ is the wavevector of the Dirac point in monolayer graphene and $t_\theta$ is the modulus of the amplitude of the main Fourier components of the interlayer potential. It is thus worth checking the validity of this formula for the whole range of angles studied here, since it would then be straightforward to extract $v_F$ and the strength of the interlayer interaction directly from our dataset. The DFT results together with two sets of TB calculations are displayed in Fig 2(c). The TB sets differ only in the first-neighbour in plane hopping parameter to get $v_{F1} = 1.1 \cdot 10^6$ m/s, similar to the value of continuum model, and a reduced $v_{F2} = 0.8 \cdot 10^6$ m/s, similar to the one obtained by DFT. All calculations show that it is possible to recover $v_F$ from the slope of $\Delta E_{VHs}$ *vs* $\sin(\theta/2)$ as predicted by the continuum model [40]. To simulate an increase of the interlayer coupling, DFT calculations have also been performed by decreasing by 0.32 Å the equilibrium interlayer distance obtained from the DFT calculation. As can be seen in Fig. 2(c), the only consequence is a rigid decrease of $\Delta E_{VHS}$, which corresponds to a larger value of $t_\theta$, in agreement with Eq. (1). Finally, figure 2(d) shows that our experimental results for the $\Delta E_{VHs}$ also conform to the theoretical predictions. Therefore, fitting the data of Fig. 2(d) with expression (1) allows deriving an experimental value of $t_\theta=0.108$ eV valid in the range 1°<θ<10°and similar to the theoretical one [9, 13, 17] and a value $v_F=1.12 \cdot 10^6$ m/s, consistent with what was reported previously for monolayer like graphene on SiC-C face [24, 31, 32, 36].

As shown, the position of vHs is very sensitive to the interlayer distance, which makes crucial the incorporation of vdW interactions in the DFT calculations. We find the DOS to be essentially unperturbed by the spatial modulation of the interlayer distance with MP periodicity, since the DOS obtained for fully relaxed layers, and thus corrugated up to 0.1 Å as shown below, is essentially the same as the one obtained when both layers are forced to remain flat [40].

We have investigated the robustness of vHs against perturbations due to subsurface stacking. In our experiments, we have taken STS data on multilayer stacks (> 5 layers thick), and considered only the rotation between the two upmost surface graphene layers. Thus, we believe it is important to assess experimentally the role played by deeper layers. This can be done here by investigating domains showing multiple MP which reveals the existence of two consecutive rotations, by small angles, between the three upmost layers [25, 27]. Spectra taken on multiple MP with two very different values of $\theta$ are shown in figures 3(a) and 3(b) respectively. In Fig. 3(a) one identifies a multiple MP with 2 superstructures of 1.52 nm and 3.7 nm periodicity. The former corresponds to a rotations of $\theta=9.3°$ between the first and second graphene layers [46]. Comparing the spectrum in Fig. 3(a) to the one shown in Fig. 1(d) for a simple MP with $\theta=9.6°$, we find that both the line shape and the vHs separation (about 1.8 V) are similar. In Fig. 3(b) the multiple MP shows two periods with values 3.0 and 12.7 nm, the latter corresponding to a rotation of $\theta=1.1°$ between the first and second graphene layers. Local spectra reveal two peaks very close to zero bias, whose intensity strongly varies with position inside the large MP period, the same as found for the simple MP with $\theta=1.4°$ of Fig. 1. Figure 1(e) finally shows that $\Delta E_{VHs}$ in STS spectra of multiple MP (empty circles) are essentially the same as for simple MP (filled circles) when considering only the rotation $\theta$ between the two upmost layers.

The question of the apparent corrugation of the MP in STM images is an important issue, since the corrugation of the MP has been considered as an indicator of the strength of the interlayer interaction and as an evidence for structural differences [13, 36, 39] which could explain the conflicting results reported for the existence of vHs in twisted multilayer samples from different origins. Figure 4(a) and 4(b) show representative STM images for two simple MP corresponding to 6.1° and 3.3° rotations angles respectively, which present then very different corrugations. The MP corrugation, defined here as the height difference between the highest and lowest areas along the MP axis, has been measured on more than 20 different MP, using different tips, tunneling impedances (from 80 MΩ to 5 GΩ) and sample biases (from 10 mV to 1 V). For a given MP, the measured corrugation depends on the tip state and tunneling parameters, only the maximum and minimum values of the measured corrugation were included to construct the graphs in Fig. 4(c). It appears immediately that the corrugation tends to increase with decreasing angle and that the contrast varies considerably (up to a factor of 3) with the experimental conditions for low angles (<4°). This gives evidence for a significant electronic effect in the contrast of large MP as previously reported [26, 27, 47, 48].

Comparison with DFT computed corrugations allow a quantitative estimate of the relative importance of topographic and electronic effects in the apparent contrast found by STM. Figure 4(d) shows the theoretical STM image for a MP with 6.01°. Ab-initio calculation reproduces quite nicely the experimental MP corrugation measured for the $\theta= 6.1°$ MP of Fig 4(a) (similar agreement is obtained for other angles). The bright areas correspond to a local AA stacking, as quoted previously [26, 47, 48]. As shown in Figure 4(e), where the interlayer distance (dots) and the total computed STM corrugation (lines) are plotted, only half of the total corrugation (0.14 Å) is of topographic origin (0.075 Å). Interestingly, for this MP, the interlayer distance varies between values close to those computed for infinite AA and AB graphene bilayers (red and blue lines in Fig. 4(e)). Thus, although the regions with e. g. AA-like stacking are quite small for this $\theta= 6.01°$ value, typically 7 C hexagons in size [47], the interaction between planes is sufficient as to set the interlayer distance close to the value found for the uniform AA stacked layers. Thus, despite the huge apparent corrugation measured by STM in large MP, our calculations set the upper limit to the actual topographic corrugation of twisted graphene layers to be 0.11 Å [40], the difference in interlayer distance between the uniform AA (3.495 Å) and AB (3.384 Å) stacked phases.

We conclude from this joint experimental and theoretical work, that rotating two graphene layers between 1º and 10º gives rise to logarithmic vHs. Relevant information about their properties can be extracted from the angular dependence of the vHs energy separation. In addition, our data reveal that they are robust against perturbations and deviations from the ideal picture, a key point to exploit the singularity of twisted graphene layers in real systems.

Acknowledgments: We thank J. M. Soler, F. Hiebel, D. Mayou and V. Olévano for fruitful discussions. This work was supported by Spain's MICINN under Grants No. MAT2010-14902, No. CSD2010-00024, and No. CSD2007-00050, and by Comunidad de Madrid under Grant No. S2009/MAT-1467. M.M.U., I.B., P.M, J.Y.V., LM and J.M.G.R. also acknowledge the PHC Picasso program for financial support (project No. 22885NH). I.B. was supported by a Ramón y Cajal project of the Spanish MEC. L. M., P. M. and J. Y. V. acknowledge support from Fondation Nanosciences (Dispograph project).

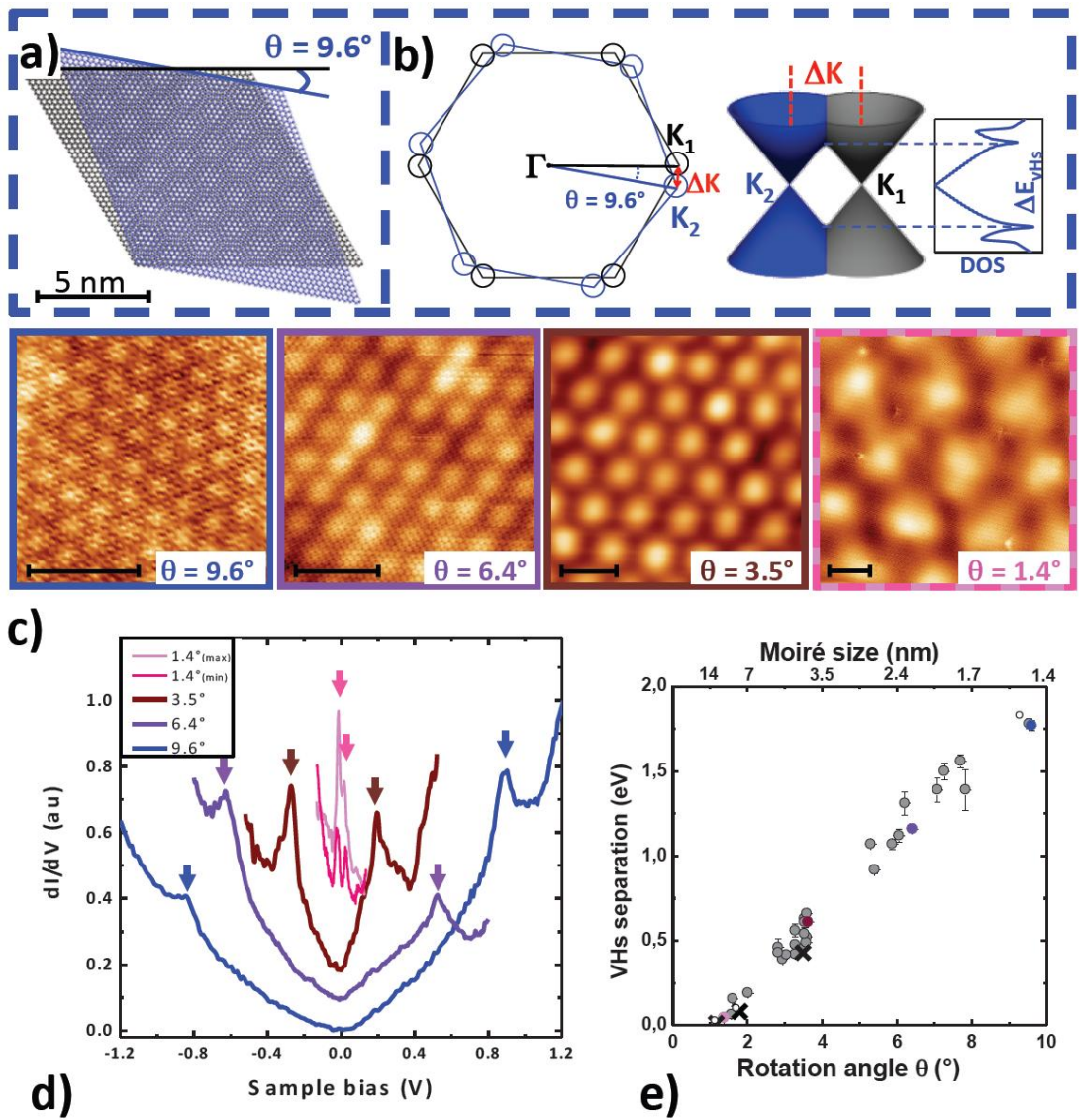

**Figure 1:** (Color online) (a) Illustration of a MP arising from a rotation angle θ=9.6°. (b) Emergence of vHs as a consequence of the rotation in reciprocal space. (c) STM images of several MP with different θ. The scale bar is 5.0 nm. (d) LDOS spectra taken on the MP shown in (c). The curves are shifted vertically for clarity. The arrows point the vHs. For θ=1.4°, max (min) indicates a spectrum taken on a bright (dark) area. (e) vHs separation as a function of θ. Crosses: data from Ref. [Andrei]. Filled and open circles: this work. Colored dots refer to the spectra displayed in (d) (same color code). Open circles correspond to multiple MP discussed in Figure 3.

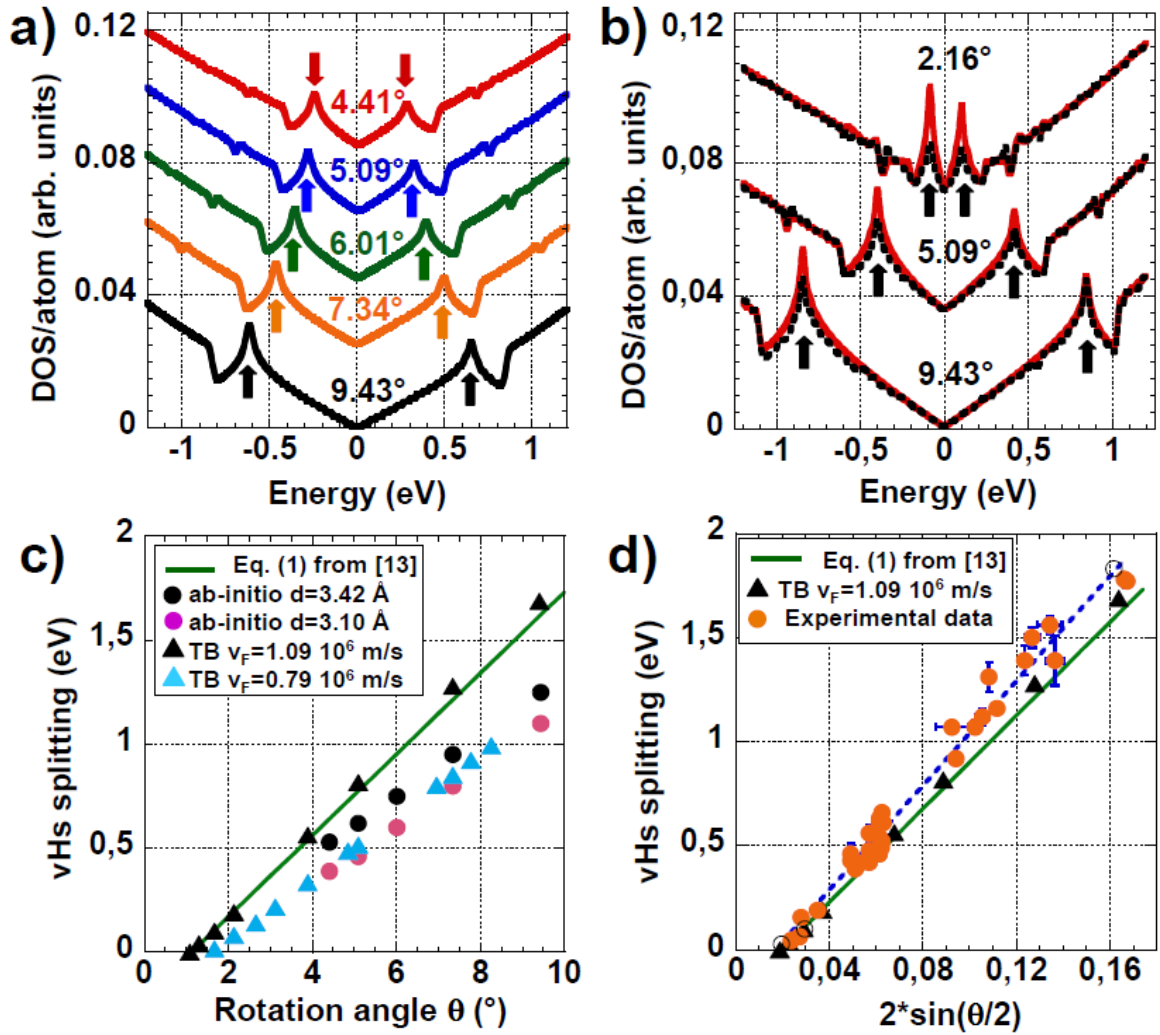

**Figure 2:** (Color online) (a) Ab-initio calculations of the total DOS, interlayer distance d=3.42 Å. (b) DOS calculations using the TB formalism with $v_F$ = 1.1 $10^6$ m/s. The red continuous line is the LDOS in AA stacked areas, the black dotted line is the total DOS. The spectra have been shifted for clarity. (c) Computed vHs splitting as a function of θ. Green line: Eq. (1) with $v_F$=1.0 $10^6$ m/s and $t_\theta$=0.11 eV [13], black (pink/gray) circles: ab-initio calculation with d=3.42 Å (d=3.10 Å), black (blue/grey) triangles: TB calculations with $v_F$ = 1.09 $10^6$ m/s ($v_F$ = 0.79 $10^6$ m/s). (d) Filled orange/grey circles: experimental data as in Fig. 1(c) compared to some calculations displayed in (c) with the same color code. The dotted (blue) line is a fit of the experimental data using Eq. (1) with $v_F$=1.12 $10^6$ m/s and $t_\theta$=0.108 eV.

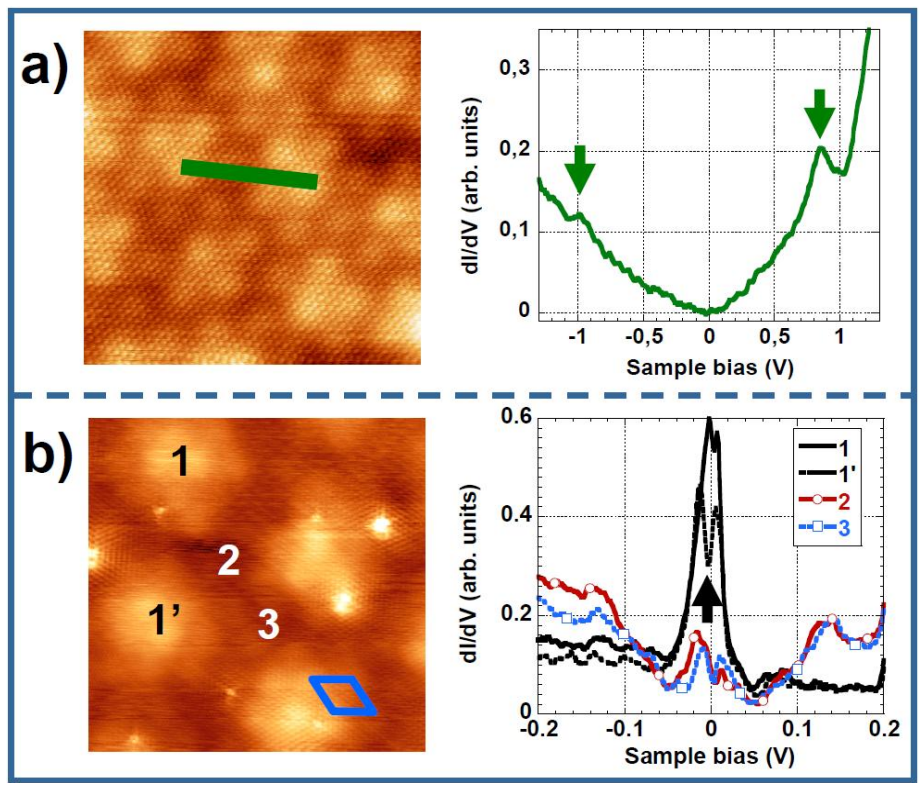

**Figure. 3**: (Color online) (a) Left: 10.6x10.6 nm² STM image of a MP with 2 periods: 1.52 nm and 3.7 nm. The smaller period corresponds to the MP between 1st and 2nd surface layers. Right: Average spectrum taken on the green line. (b) Left: 24x24 nm² STM image of a MP with 2 periods: 12.7 nm and 3.0 nm (blue diamond). The larger period corresponds to the MP between 1st and 2nd surface layers. Right: Local spectra taken on the spots labelled on the image. Arrows point the vHs.

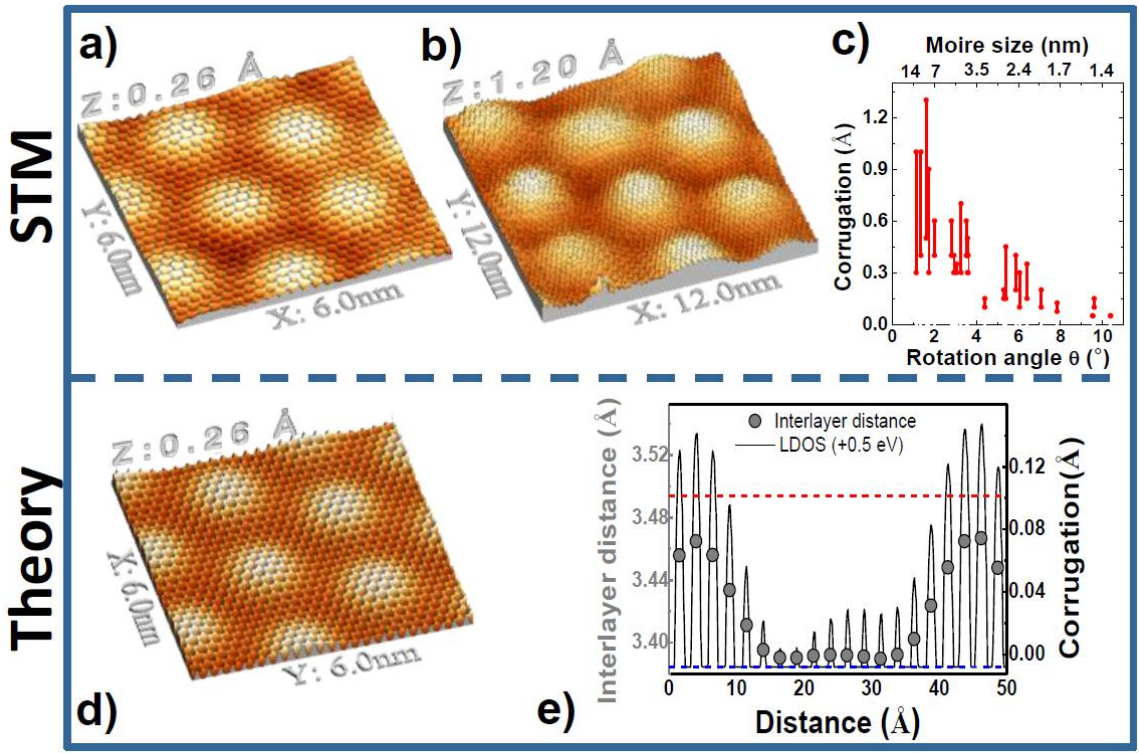

Figure 4 (Color online): (a) STM image at sample bias V=+0.5 V of a MP with P=2.32 nm (θ=6.08°). The measured corrugation is 0.14 Å. (b) STM image at V=+0.2 V of a MP with period P=4.3 nm (θ=3.28°). (c) Range of STM measured MP corrugation as a function of rotation angle θ. (d) Simulated STM image for a bilayer rotated θ=6.01° at +0.5 eV. e) Atomic (dots) and total (line) corrugation for the bilayer image shown in (d) along the long MP diagonal. The blue and red horizontal dashed lines indicate the interlayer spacing computed for AB and AA bilayers respectively.

# Supplementary Material for

# Unravelling the intrinsic and robust nature of van Hove singularities in twisted bilayer graphene

I. Brihuega, P. Mallet, H. González-Herrero, G. Trambly de Laissardière, M. M. Ugeda, L. Magaud, J.M. Gómez-Rodríguez, F. Ynduráin, and J.-Y. Veuillen(*)

It consists in 5 sections, labelled from I to V, devoted to:
-**section I**: sample overview,
-**section II**: spatial variation of the vHs amplitude in small period MP,
-**section III**: details on ab-initio calculations of LDOS and band structure of twisted bilayers,
-**section IV**: angular variations of the vHS splitting using different computation schemes,
-**section V**: details on the calculation of the moiré corrugation.

I. Sample overview:

A large scale image of the sample is shown in figure S1. It shows a wide terrace, divided by wrinkles (labelled W) and by lines of beads (indicated by arrows) into coherent domains with a typical size of 100 nm. The height of the wrinkles is in the nanometre range (1-3 nm). The overall morphology of the sample is similar to the one reported in Ref. [1] for a sample of the same origin. The moiré patterns (MP) discussed in the main text are found within the coherent domains.

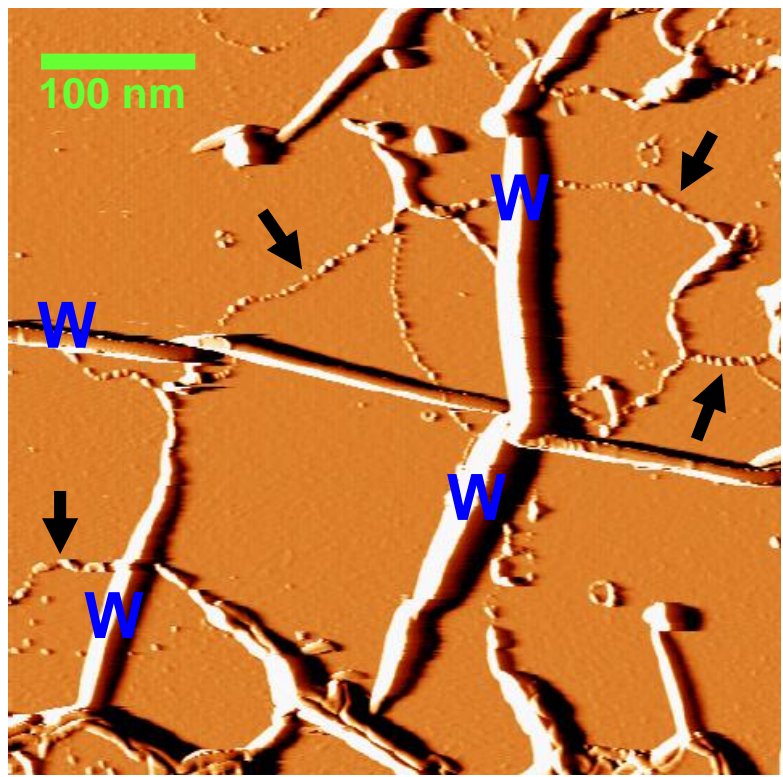

*Figure S1*: Large scale image of the sample.
*Image size: 500x500 nm², Sample bias Vs= 3.14 V, tunnelling current I=0.5 nA.*
*A derivative filter has been applied on the image to reveal the details of the structure.*

The domains size in our samples are somehow smaller than for samples produced by optimized growth techniques [2], which has some advantages for the experimental work we present in the main paper since it allows probing different spots on neighbouring coherent domains with the same tip. Hence we could verify that i) for a coherent domain the vHs show up in the spectra everywhere, from the centre up to the boundary, with essentially the same splitting and ii) the peaks in Fig. 1(d) are perfectly reproducible and not due to tip artefacts as confirmed by interleaved spectra taken on the two sides of the boundaries of the MP.

The sample thickness estimated from Auger spectroscopy is at least 5 graphene layers. Thus the first and second surface graphene layers should be almost neutral. Indeed, no significant charge transfer from the substrate could be detected beyond 3 layers at the graphene-SiC interface [3, 4]. Thus the Dirac point is very close to the Fermi level at the surface of our sample. As a result, the positions of the LDOS singularities at positive and negative biases are almost symmetric with respect to the Fermi level in the spectra of Fig. 1. Electrostatic fields effects such as the ones reported in Ref. [5] are thus not important for the purpose of our work (they should however show up for thinner samples).

II. Spatial variation of the vHs amplitude in small period MP:

Figures 1(d) and 3(b) of the main paper show that for large period (small rotation angle) MP the amplitudes of the LDOS peaks corresponding to the vHs are strongly modulated within the moiré (pseudo) unit cell. This is consistent with previous results [6]. In Ref. [6] it was shown that this tendency towards LDOS localization was suppressed for larger rotation angles ($\theta=3.5°$). Figure S2 below shows that the same happens in our sample.

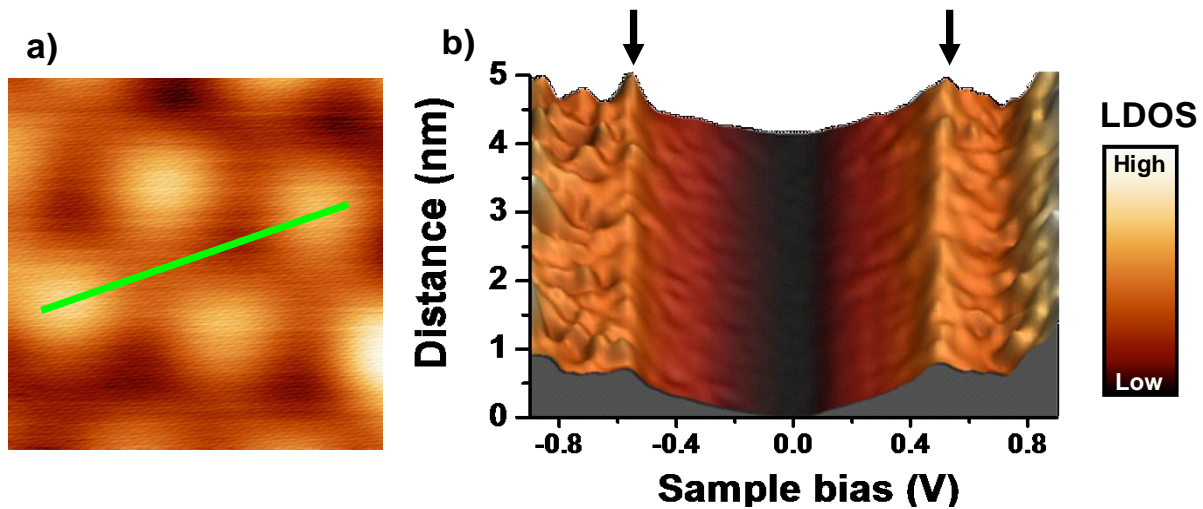

*Figure S2:* Homogeneity of the electronic structure on a small period MP.
(a) STM image on a MP of period 2.66±0.2 nm ($\theta\approx5.30°$). Image size: 6.0x6.0 nm²,
Vs=+0.50 V, I=0.47 nA. 32 spectra (stab: -1.09 V, I= 1.7 nA) were taken along the long diagonal of the MP (indicated by the green line). (b) Map of conductance (dI/dV) spectra along the green line in a).

A set of spectra was taken along the green line in Figure S2(a) and the related conductance map, which gives essentially the LDOS, is shown in figure S2(b). The vHs are pointed by arrows in figure S2(b). Their amplitude is essentially constant along the long diagonal of the MP. Hence the localization of the vHs does not occur for this MP with P=2.66 nm, $\theta=5.3°$, at variance with the situation for long period MP's (figures 1 and 3). This is in agreement with calculations in figure 2(b) (main paper) which show a strong reduction of the amplitude of the vHs spatial modulations with increasing angle.

## III. Details on ab-initio calculations of DOS and band structure of twisted bilayers.

We have performed density functional calculations using the SIESTA code [7] that uses localized orbitals as basis functions. We use a double $\zeta$ basis set, non-local norm conserving pseudopotentials and for exchange and correlation we use the functional proposed by Dion et al. [8, 9] intended to include van der Waals interactions. The calculations have being performed with stringent criteria in the electronic structure convergence (down to $10^{-5}$ in the density matrix), Brillouin zone sampling (up to 2500 k-points), real space grid (energy cut-off of 500 Ryd) and equilibrium geometry (residual forces lower than $2\times10^{-2}$ eV/A). In the equilibrium geometry calculations one layer is kept fixed and planar whereas the atoms of the other layer are free to relax till the equilibrium geometry is obtained. The average distance between the two layer ranges between 3.494 Å (value for the uniform AA stacking) and 3.384 Å (value for the uniform AB stacking) depending on the MP considered. Additional atomic corrugation is found being indeed MP dependent. In this section only the results for the $\theta=6.08°$ MP are reported.

In figure S3 we show the band structure and DOS for the (5,6) ($\theta=6.01°$) MP with two basis sets wavefunctions. The equilibrium geometry is, indeed, the geometry obtained with double $\zeta$ basis. We immediately notice the extra structure induced by the interaction between the two layers. The logarithmic vHs are apparent and do not seem to depend much on the basis set used.

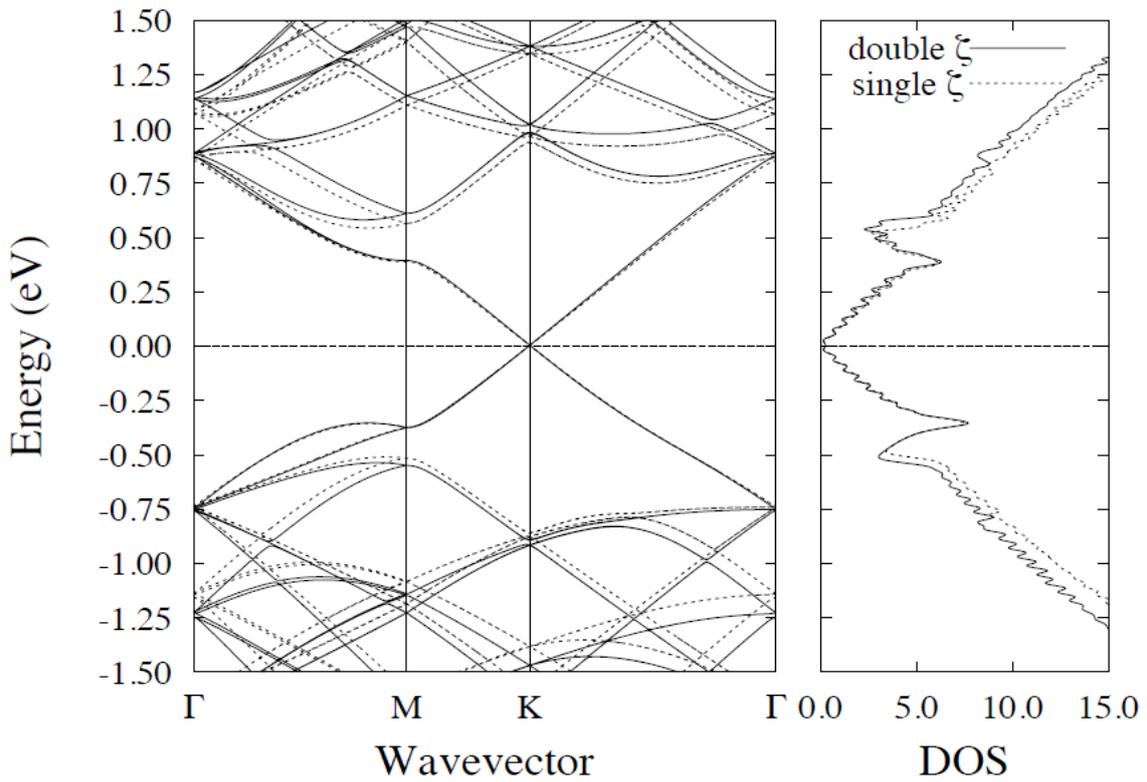

*Figure S3:* Band structure and DOS of the (5,6) MP calculated with two different basis functions sets. The equilibrium geometry is the one obtained with the more complete basis set (double $\zeta$ basis).

To study the effect of the atomic corrugation on the electronic structure we have performed the calculation for two different atomic geometries, namely, the equilibrium geometry and two planar graphene layers separated by the average distance between the two layers in the equilibrium geometry. The results are shown in figure S4. We observe that the corrugation of the atoms scarcely affects the electronic structure and in particular the vHs singularities.

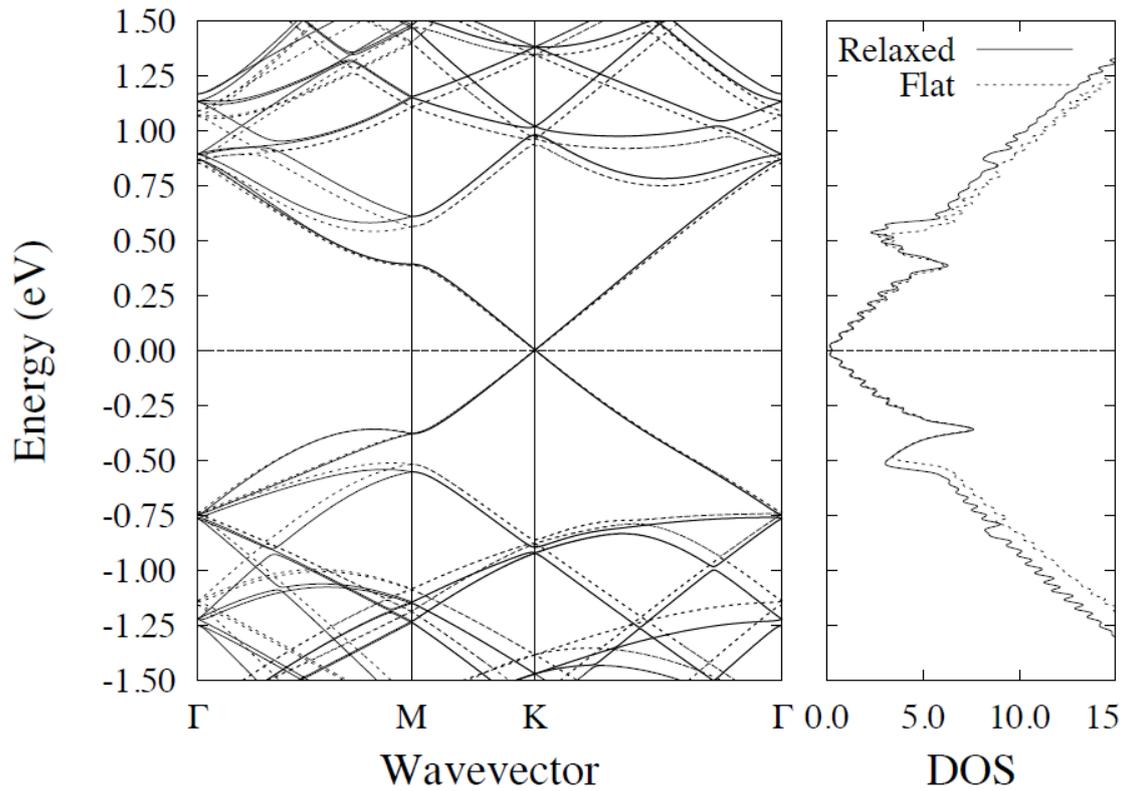

*Figure S4:* Calculated electronic structure (band structure and DOS) of the (5,6) MP considering the relaxed structure (heavy lines) and a flat layer (broken lines). In both cases a double $\zeta$ basis is being considered.

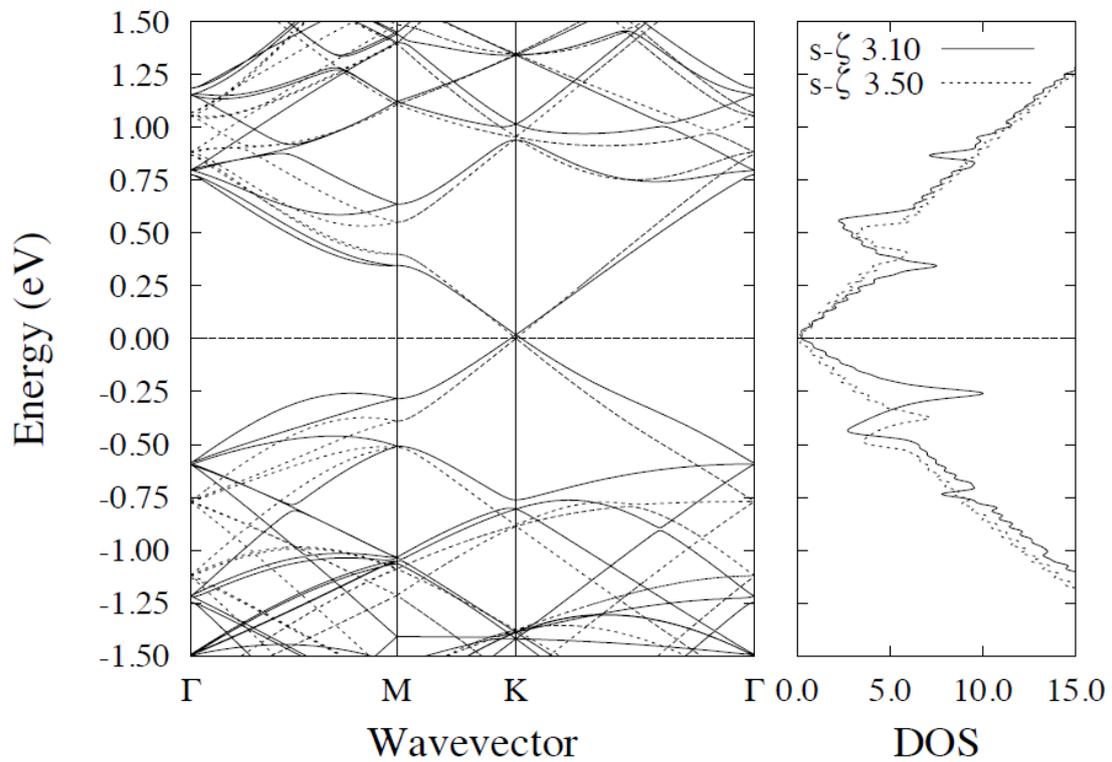

*Figure S5.* Band structure and DOS of the (5,6) MP calculated for two planar layer separated by two distances. Heavy and broken lines stand for layers separated by 3.10 Å and 3.50 Å respectively.

In order to evaluate the effect of the inter layer separation on the electronic structure and in particular in the vHs, we have calculated, using a single $\zeta$ basis, the band structure and densities of states for two different separations between two planar layers. The results are shown in figure S5. The results indicate how sensible the vHs are with respect to the interlayer separations.

Finally, we consider the importance of the van der Waals interaction in the electronic structure. To this purpose we present in figure S6 the calculated electronic structure considering the functional used throughout this work [8] and a conventional GGA functional that does not include at all van der Waals interaction. The calculations are for the equilibrium geometry obtained above. We find that, at least for energies close to the Dirac point, the electronic structure in both cases is almost identical. It is worth to indicate that whereas the geometry and layer separation is mainly governed van der Waals forces the electronic structure is independent of the functional used.

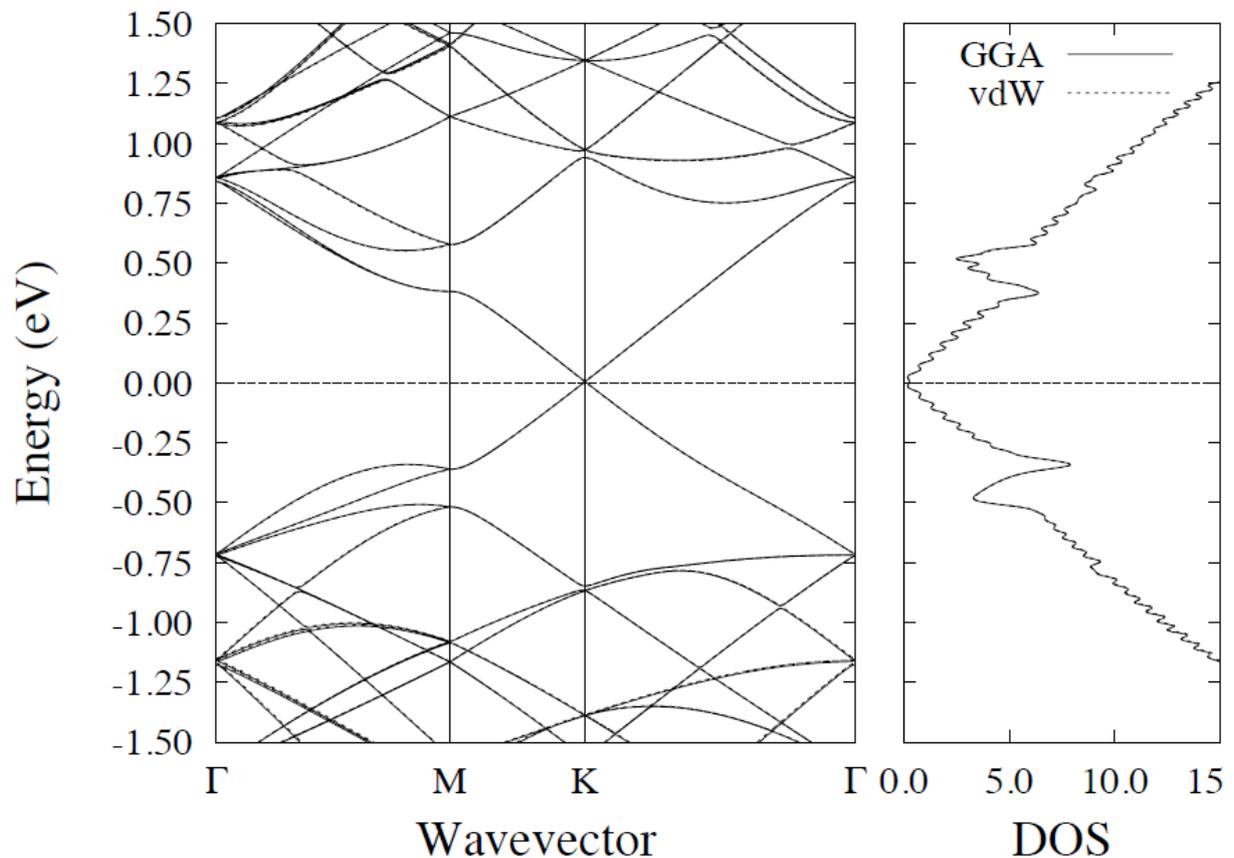

*Figure S6*: Band structure and DOS of the (5,6) MP calculated with two different functionals. Heavy line stands for the results obtained with a conventional GGA functional and the broken line (almost not seen due to the overlap with the heavy line) corresponds to the results obtained with the Dion et al. [8] functional including the van der Waals interaction.

## IV. Angular variations of the vHS splitting using different computation schemes.

In figure 2(c) of the main paper we have displayed the variation with the rotation angle $\theta$ of the vHs splitting $\Delta E_{VHs}$ computed with the various theoretical approaches used in this work. Here we demonstrate that all the values $\Delta E_{VHs}(\theta)$ obtained by the different theoretical approaches are well described by the simple equation [6]:

$$\Delta E_{VHs} = 2\hbar \cdot v_F \cdot \Gamma K \cdot \sin(\theta/2) - 2t_\theta, \qquad (1)$$

where $v_F$ is the Fermi velocity *computed* with the same techniques for *monolayer* graphene, $\Gamma K = 1.703$ Å$^{-1}$ is the wavevector of the Dirac point in monolayer graphene and $t_\theta$ is the modulus of the amplitude of the main Fourier components of the interlayer potential. We moreover show that changes in the computational parameters lead to physically reasonable changes in the $v_F$ and $t_\theta$ values deduced from equation (1).

Figure S7 displays the angular variation of the vHs splitting $\Delta E_{VHs}(\sin(\theta/2))$ obtained in our different simulations, together with a simplified representation of the experimental data (it is actually the fit of the STS data with Eq. (1) given in Fig. 2(d)). The linear fit of the computed vHs splitting with Eq. (1) is excellent, as shown in Fig. S7. The fitting parameters are listed in Table S8. Table S8 shows that the values of $v_F$ deduced from the slope of $\Delta E_{VHs}(\sin(\theta/2))$ are in good agreement with the Fermi velocities computed for *monolayer* graphene using the same simulation techniques.

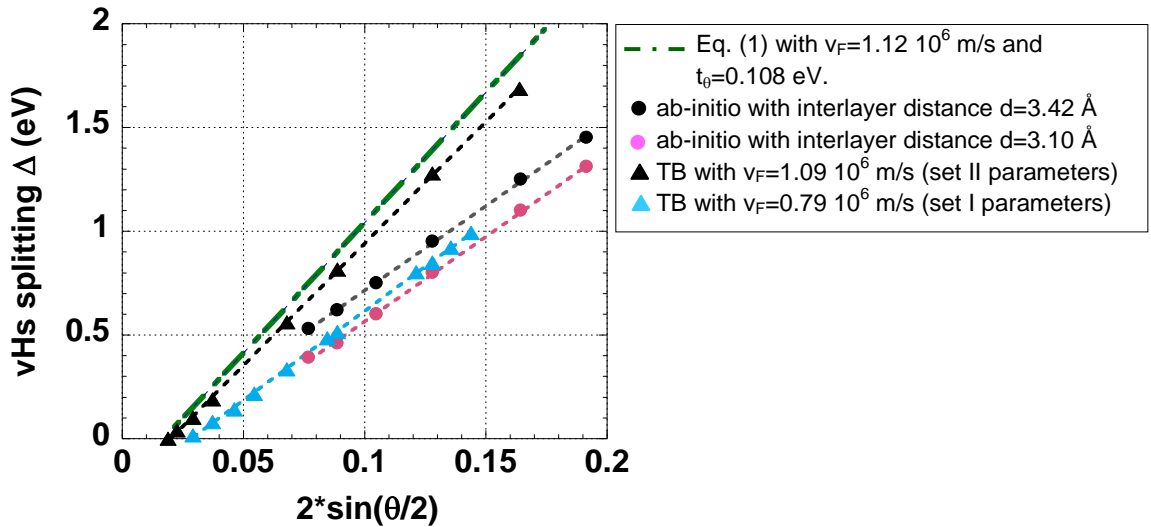

*Figure S7*: Linear fit of the theoretical vHs splitting $\Delta E_{VHs}(\sin(\theta/2))$ for $\theta \leq 11°$ and schematics of experimental data. *Green dot-dashed line:* Linear fit of the experimental results (Fig. 2(d)) with equation (1) with $v_F=1.12\ 10^6$ m/s and $t_\theta=0.108$ eV. Black (blue) triangles: TB results with set II (set I) parameters, black (pink) circles: ab-initio calculations (SIESTA) with interlayer distance d=3.42 Å (d=3.10 Å).

| Computation technique | Fitted $v_F$ value (m/s) | Fitted $t_\theta$ value (eV) | Computed Fermi velocity for <u>monolayer</u> graphene (m/s) |
|---|---|---|---|
| TB set I parameters | $0.77\ 10^6$ | 0.122 | $0.79\ 10^6$ |
| TB set II parameters | $1.05\ 10^6$ | 0.117 | $1.09\ 10^6$ |
| Ab-initio d=3.42 Å | $0.73\ 10^6$ | 0.048 | $0.804\ 10^6$ |
| Ab-initio d=3.10 Å | $0.73\ 10^6$ | 0.126 | $0.804\ 10^6$ |

<u>Table S8</u>: *Parameters obtained by fitting with equation (1) the values of the vHs splitting $\Delta E_{VHs}(\sin(\theta/2))$ computed using the different simulation techniques described in the main text. The last column is the value of the Fermi velocity computed for monolayer graphene using the same procedures. d is the interlayer distance in the ab-initio calculations.*

Table S8 gives some additional information:
- Only the first neighbour in plane hopping parameter changes between TB set I and TB set II (the parameterization of the interlayer interaction remains the same). As a consequence only the value of $v_F$ changes in the fit (in the same way as for monolayer graphene), while $t_\theta$ is unaffected as expected.
- Only the interlayer interaction should be affected when the distance between graphene planes is reduced in ab-initio calculations. Accordingly, we notice that only the $t_\theta$ value changes in the fit, while $v_F$ remains the same.

Thus it appears that Eq. 1 gives a good quantitative account of our theoretical results with $t_\theta \approx 0.1$ eV and using the Fermi velocity computed for monolayer graphene. As expected, this value of $t_\theta$ is consistent with (half) the width of the partial gap which develops in band structure calculations (this work, [10, 11]) at the mid-point along the line which connects the (closest) Dirac points $K_1$ and $K_2$ (cf. Fig. 1(b)) in twisted bilayers (this mid-point corresponds to the M point in the supercell Brillouin zone for the (5,6) MP decribed in section III or for the (6,7) MP of Ref. [11]).

The fact that DFT based computations underestimates the Fermi velocity is a well known effect. Taking into account the many body electron-electron interactions in calculations based on the GW formalism leads to a renormalization of the bands and restores the experimental value [12].

V. Details on the calculation of the moiré corrugation.

*V.1 Local variations of the interlayer distance (topographic corrugation) :*

In figure S9, we first illustrate the relationship between stacking and local interlayer distance for the (5,6) MP (rotation angle θ=6.01°). The topographic profile displayed in Fig. 4(e) was taken approximately along the long diagonal of the unit cell of the MP. Figure S9(a) shows that the areas with well defined AA stacking consists of only 7 graphene hexagons. This means that only ≈13% of the C atoms (48/364) rest in these areas. For θ=6° a similar fraction of AA stacking (10% to 20%) was estimated in Ref. [13].

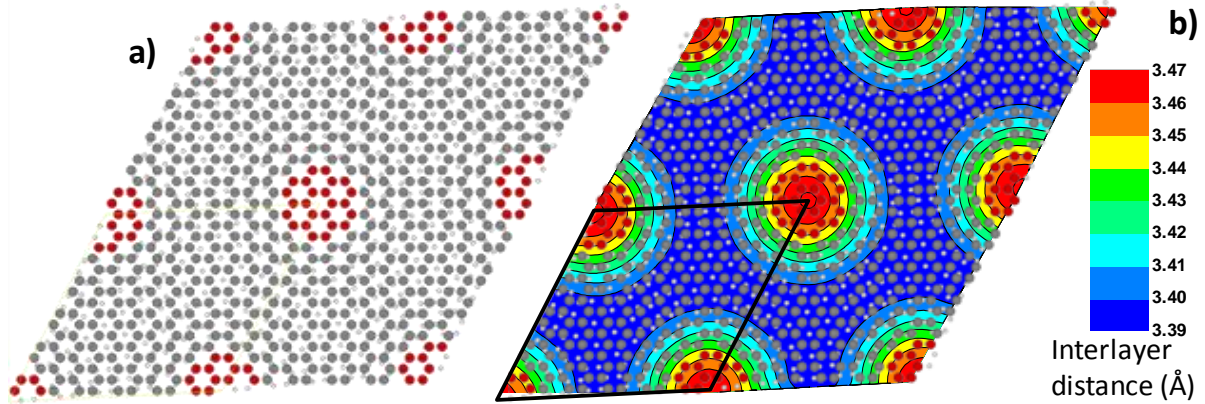

*Figure S9*: Structure of the (5,6) MP. (a) Stacking of the (5,6) bilayer. The larger circles represent the atoms of the top layer. Red dots indicate areas with approximate AA stacking. Four unit cells are represented (one is highlighted by the black line in b)). (b) Computed interlayer distance coded in colour as a function of the stacking. The atomic stacking scheme of a) is superimposed on the colour map. Areas with AA stacking (red atoms) show the largest local interlayer spacing.

*V.2 Bias (energy) dependence of the total corrugation :*

Table S10 summarizes the corrugations obtained for simulated STM images, within the Tersoff-Haman approximation at constant isosurface =0.0000125 electrons/Bohr$^3$, of the (5,6) MP including atomic relaxation. For comparison, the "topographic" corrugation (i. e. the variation of the interlayer distance) is 7.5 pm as shown on figures 4(e) and S9(b).

| Energy (eV) or tunneling bias (V) | Simulated STM Corrugation (pm) |
|---|---|
| +0.2 | 14 |
| +0.4 | 14 |
| +0.6 | 13.5 |
| +0.8 | 11 |

*Table S10*: corrugation of computed STM images of the (5,6) MP as a function of sample bias (or energy integration window above the Dirac point).

The total corrugation is larger than the topographic one, thus electronic effects contribute to the contrast of the MP in computed STM images. The contrast does not change with bias in the +0.2V/+0.6V range. Moreover, the corrugation does not depend on the tunnelling (sample) bias being larger (V=+0.6V) or smaller (V=+0.2V) than the vHs energy (E=+0.39 eV). This is consistent with the weak variation of the amplitude of the vHs in the MP (pseudo) cell found for large angles (θ>5°) (cf. Ref. [6], figure 2(b) of the main paper and section II of this supplementary material).

*V.3 Total corrugation for flat and relaxed twisted bilayers.:*

We have compared the total corrugation of computed STM images for (5,6) MP with and without local variations of the interlayer distance (denoted as "relaxed" and "flat" layers respectively below). The mean interlayer spacing (integrated over the supercell) was the same in both cases. The results are shown in figure S11.

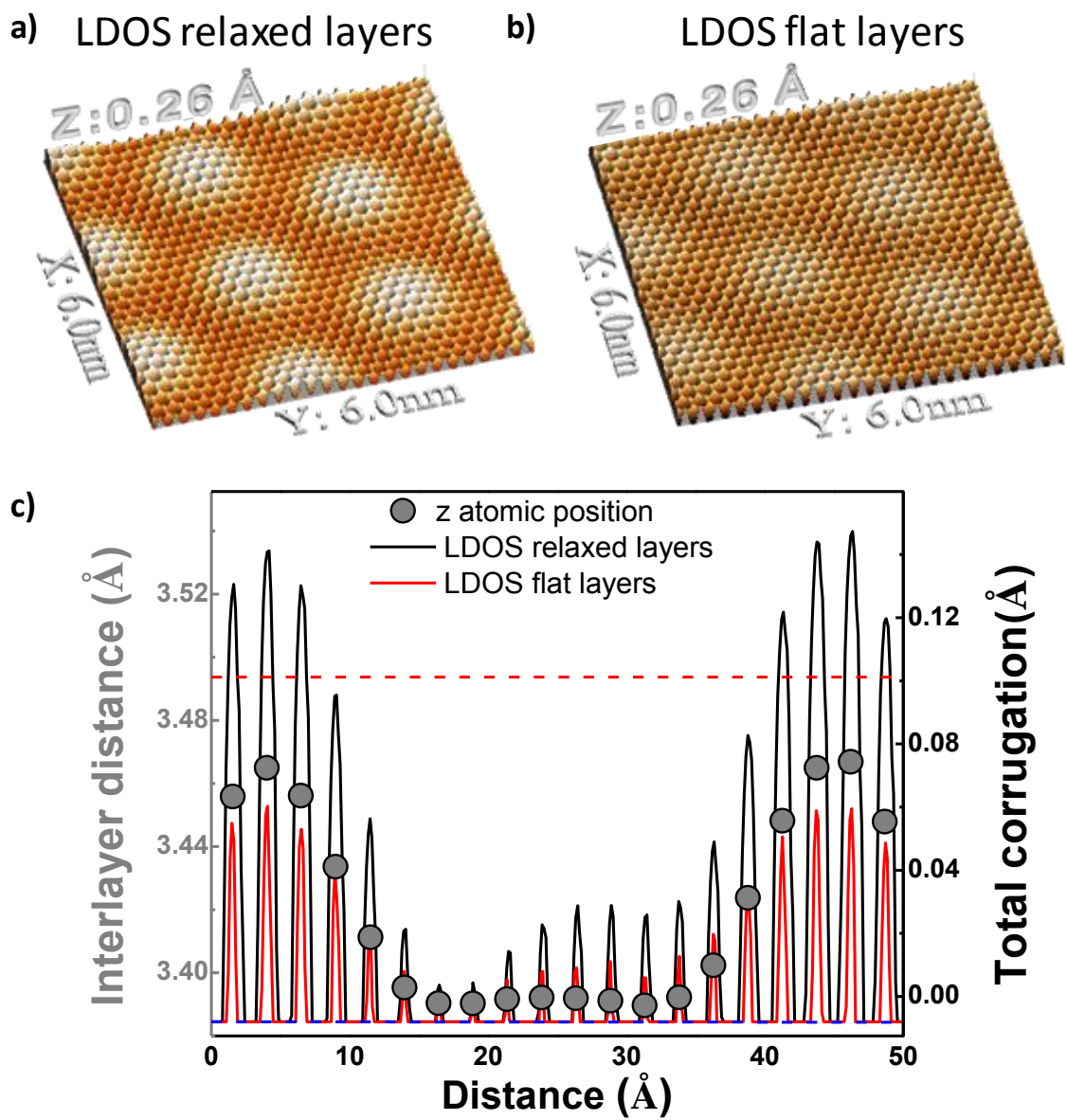

*Figure S11:* Total corrugation of the (5,6) MP computed for relaxed and flat layers. (a) STM image computed for a relaxed bilayer for tunnelling bias +0.5 V. (b) STM image computed for a flat bilayer for tunnelling bias +0.5 V. (c) Corrugation profiles along the long diagonal of the MP. Left scale: variation of the interlayer distance (related to the topographic corrugation) for the relaxed bilayer in a). Right scale: total corrugation of the images in a) (grey line) and in b) (red line). The blue and red horizontal dashed lines indicate the interlayer spacing computed for AB and AA bilayers respectively (left scale).

Figure S11(a) is the same as figure 4(d). Figure S11(b) shows a contrast related to the MP which is purely of electronic origin [14] since the layers are forced to remains flat there. The gray lines and dots in figure S11(c) are the same as in figure 4(e). The red line is the corrugation computed for the flat bilayer of figure S11(b). It turns out that this corrugation, which is only related to LDOS variations, corresponds almost exactly to the difference between the "topographic" and total corrugation for the relaxed bilayer (gray dots and lines in figure S11(c)). This means that the LDOS modulations in the MP cell are not affected by the (small) variations of the interlayer distance in the relaxed case.

*V.4 Corrugation for a bilayer with large twist angle.*

We have also computed the total corrugation (including atomic relaxation) for a (2,6) MP with a large rotation angle ($\theta$=32.2°). The results are displayed in figure S12.

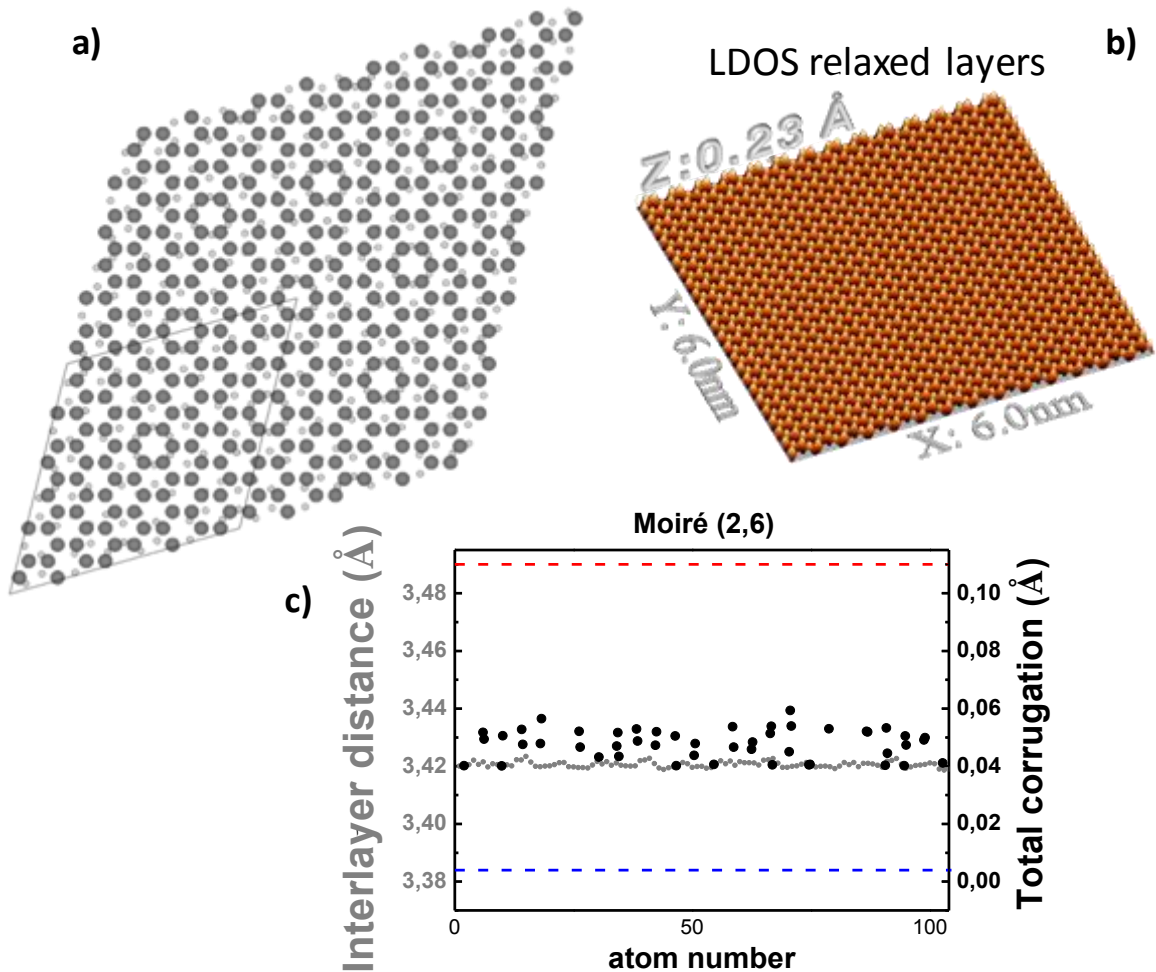

*Figure S12: Stacking and corrugation for a MP with large twist angle ($\theta$=32.2°). (a) Stacking sequence of the two layers. The larger dots indicate the C atoms of the top layer. (b) STM image computed for a relaxed bilayer for tunnelling bias +0.2 V. (c) Interlayer distance (left scale, small gray dots) and variation of the apparent height (right scale, large black dots) computed at a tunnelling bias +0.2V for all the atoms in the unit cell (thin lines in a)). The blue and red lines indicate the interlayer spacing computed for (uniformly) AB and AA stacked bilayers respectively.*

From figure S12(a) it appears than we can no longer identify areas with well defined AA stacking for this large rotation angle, in agreement with Ref. [13]. Both the interlayer distance and the apparent height are virtually the same for all the atoms in the MP cell as shown in figure S12(c) (small grey dots, left scale). Thus, as shown in Fig S12(b), the MP becomes hardly visible in STM images for large rotation angles, in agreement with previous findings [1, 13, 15, 16].

*V.5 Variation of the interlayer distance with twist angle.*

We have computed the variation of the interlayer distance in the MP by means of DFT calculations with van der Waals functional for several rotation angles. This quantity, which corresponds to the topographic corrugation of the MP is represented in Figure S13.

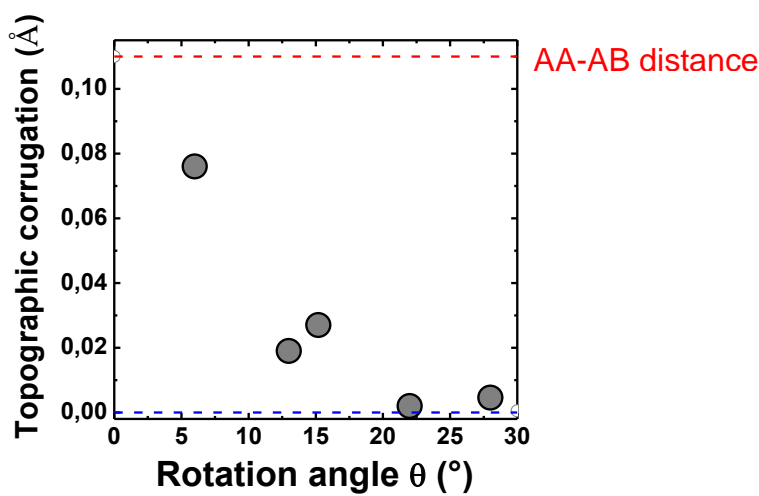

Figure S13: Variation of the interlayer distance (difference between the maximum and the minimum distances between layers over the MP unit cell) as a function of the rotation angle $\theta$. The blue dashed line is zero. The red dashed line is the difference in the interlayer distance between uniform AA and AB stacked bilayers (12 pm).

The variation of the interlayer distance shows the same trend as the experimental MP corrugation shown in Figure 4(c). It markedly decreases with increasing angle to values well below the difference between uniform AA and AB stacked bilayers (red dashed line) for $\theta \geq 15°$. This is in agreement with the argument presented in Ref. [13], where it was shown that local AA and AB stacking can hardly be defined beyond 15°. The lack of clear differentiation of the local environment of the C atoms over the MP results in very small topographical corrugation.